# Algorithms for the Problems of Length-Constrained Heaviest Segments

Md. Shafiul Alam and Asish Mukhopadhyay ⋆

School of Computer Science, University of Windsor,
401 Sunset Avenue, Windsor, ON, N9B 3P4, Canada

**Abstract.** We present algorithms for length-constrained maximum sum segment and maximum density segment problems, in particular, and the problem of finding length-constrained heaviest segments, in general, for a sequence of real numbers. Given a sequence of $n$ real numbers and two real parameters $L$ and $U$ ($L \leqslant U$), the maximum sum segment problem is to find a consecutive subsequence, called a segment, of length at least $L$ and at most $U$ such that the sum of the numbers in the subsequence is maximum. The maximum density segment problem is to find a segment of length at least $L$ and at most $U$ such that the density of the numbers in the subsequence is the maximum. For the first problem with non-uniform width there is an algorithm with time and space complexities in $O(n)$. We present an algorithm with time complexity in $O(n)$ and space complexity in $O(U)$. For the second problem with non-uniform width there is a combinatorial solution with time complexity in $O(n)$ and space complexity in $O(U)$. We present a simple geometric algorithm with the same time and space complexities.

We extend our algorithms to respectively solve the length-constrained $k$ maximum sum segments problem in $O(n + k)$ time and $O(\max\{U, k\})$ space, and the length-constrained $k$ maximum density segments problem in $O(n \min\{k, U - L\})$ time and $O(U + k)$ space. We present extensions of our algorithms to find all the length-constrained segments having user specified sum and density in $O(n + m)$ and $O(n \log(U - L) + m)$ times respectively, where $m$ is the number of output. Previously, there was no known algorithm with non-trivial result for these problems. We indicate the extensions of our algorithms to higher dimensions. All the algorithms can be extended in a straight forward way to solve the problems with non-uniform width and non-uniform weight.

The algorithms have potential applications in different areas of biomolecular sequence analysis including finding CG-rich regions, TA and CG-deficient regions, CpG islands and regions rich in periodical three-base patterns, post processing sequence alignment, annotating multiple sequence alignments, and computing length constrained ungapped local alignment. They also have applications in other areas such as pattern recognition, digital image processing and data mining.

---

⋆ Research supported by an NSERC discovery grant awarded to this author.

## 1 Introduction

In this paper we study two problems concerning the determination of length-constrained heaviest segments in a sequence of real numbers. The problems in their basic form are described below:

**Definition 1.** *Given a sequence of pairs of real numbers* $(a_i, w_i)$, $i = 1, 2, \ldots, n$, *with* $w_i > 0$, *and another pair of real numbers* $L \leqslant U$. *(a) The maximum sum segment problem is to find a subsequence* $a_i, a_{i+1}, \ldots, a_j$ *whose sum* $a_i + a_{i+1} + \ldots + a_j$ *is the maximum under the constraint that* $L \leqslant w_i + w_{i+1} + \ldots + w_j \leqslant U$. *(b) The maximum density segment problem is to find a subsequence* $a_i, a_{i+1}, \ldots, a_j$ *whose density* $(a_i + a_{i+1} + \ldots + a_j)/(w_i + w_{i+1} + \ldots + w_j)$ *is the maximum under the constraint that* $L \leqslant w_i + w_{i+1} + \ldots + w_j \leqslant U$.

The maximum sum segment problem, with uniform width ($w_i = 1$ for all $i$) and no restriction on segment length, was formulated by Ulf Grenander [22, 6]. He found it in the area of pattern recognition in digitized image. The original problem, as proposed by him, was in 2-dimensions. In that setting the maximum sum subarray was the estimator for the maximum likelihood of a pattern in digital image. He also simplified the problem to 1-dimension. The problem also has applications in other areas such as graphics, data mining [1, 17, 18] and bioinformatics [4]. An optimal linear time algorithm for the problem proposed by Jay Kadane is described by Bentley [6] and Gries [23]. Its space complexity is $O(1)$. The two dimensional version of the problem is to find a connected rectangular submatrix of maximum sum from a two-dimensional rectangular input matrix of real numbers [6]. Here the widths are uniform, i.e., $w_i = 1$ for all $i$, and there is no restriction on the size of the submatrix. The problem has been extended to higher dimensions [38]. In higher dimensions the problem is called the maximum sum subarray problem. The higher dimensional problem has applications in the area of data mining (dimensions being less than 4) and Monte-Carlo simulation (dimensions being high) [38]. It can be solved by reducing it to 1-dimensional problems [7, 38]. For 2-dimensional $m \times n$ matrix there are $O(m^2)$ column intervals. Each of them is solved using Kadane's linear time algorithm for maximum sum segment problem. Hence, its time complexity is $O(m^2 n)$ [7, 38]. For this case, i.e., 2-dimensions with uniform width and no restriction on length, there is a better algorithm based on distnace matrix multiplication technique [38, 37]. Its running time is subcubic.

Huang [25] introduced the restriction of length cut off $L$ in the setting of biomolecular sequence analysis to avoid reporting extremely short segments. He gave a linear time algorithm for computing the maximum sum segment of length at least $L$, but no restriction on the upper bound of its length, i.e., $U = n$. He had observed that the segments reported by the algorithm are usually much larger than $L$. From this observation Lin *et al.* [28] argued that the segments reported by the method may contain some poor and irrelevant segments. To avoid this they introduced the restriction of upper bound $U$ on the length of the segment. They proposed a linear time algorithm for the problem when there is only the upper bound $U$ on the length of the segment, but no lower bound, i.e., $L = 0$. They combined that algorithm with Huang's [25] technique to develop a linear time algorithm for arbitrary $L$ and $U$. Its space requirement is also linear. In this paper, we present an algorithm for this general problem with time complexity in $O(n)$ and space complexity in $O(U)$. We indicate the extension of this algorithm to solve the problem in higher dimension by using the technique of reducing the problem to 1-dimension [7, 38].

The $k$ maximum sum segments problem was introduced by Bae and Takaoka [5]. There was no restriction on the segment length. A natural extension of this problem is the $k$ length-constrained maximal sum segments problem. The problem is defined as follows:

**Definition 2.** *Given a sequence of real numbers $a_i$, $i = 1, 2, \ldots, n$, a pair of real numbers $L \leqslant U$ and an integer $k$ such that $1 \leqslant k \leqslant (n - U + 1)(U - L + 1) + \frac{1}{2}(U - L)(U - L + 1)$. The $k$ length-constrained maximum sum segments problem is to find $k$ subsequence of consecutive elements of length at least $L$ and at most $U$ such that their sums are the $k$ largest among all the possible segments of length at least $L$ and at most $U$.*

When there is no restriction on segment length, i.e., $L = 0$ and $U = n$, Brodal and Jorgensen [9] gave an optimal linear time, i.e., $O(n + k)$ time, algorithm for this. Their algorithm constructs a partially persistent [13] binary maximum heap that implicitly contains all the $\binom{n}{2} + n$ number of sums for all possible segments in $O(n)$ time. The heap is a modified version of the self-adjusting heap of Sleator and Tarjan [34]. The $k$ maximum sums are selected from the heap using linear time heap selection algorithm of Frederickson [16]. Brodal and Jorgensen [9] extended their algorithm to higher dimension by using the technique of reducing the problem to 1-dimension [7, 38].

Combining with their technique we extend our algorithm for the maximum sum segment problem to solve the $k$ length-constrained (i.e., arbitrary $L$ and $U$) maximum sum segments problem. Its time and space complexities are $O(n + k)$ and $O(U + k)$ respectively. Previously, there was no known algorithm with non-trivial result for this case.

For the maximum density segment problem, when the widths are uniform and there is no restriction on the segment length the maximum element in the sequence will be the solution and it can be found in a straight forward way in $n - 1$ comparisons and $O(1)$ space. When $U = L$ the problem is trivially solvable in $O(n)$ time since there are $n - U + 1$ feasible segments. When the widths are uniform, $U \neq L$ and no upper bound ($U \geq n - L$) Huang [25] showed that the length of the maximum density segment is at most $2L - 1$. So, this case is equivalent to the case when $U = 2L - 1$ and can be solved in $O(nL)$ using brute force method since the number of feasible segments is $O(nL)$. For this case Lin *et al.* [28] gave an $O(n \log L)$ time algorithm by using a method of right skew decomposition of the sequence. When the widths are uniform, and $U$ and $L$ are arbitrary Goldwasser *et al.* [20] gave an $O(n)$ time algorithm. For the general case, where the widths are not uniform and $U$ and $L$ are arbitrary, Goldwasser *et al.* [21] extended the right skew decomposition method of Lin *et al.* [28] to develop a $O(n)$-time and space algorithm. A combinatorial solution with time-complexity in $O(n)$ and space complexity in $O(U)$ was proposed by [11]. The algorithm works in an online manner. In the same paper it was pointed out that the linearity claim of a geometric approach by Kim [27] is flawed. In this paper we modify Kim's algorithm to address the flaw, while retaining the simplicity, elegance and linearity of his geometric approach. Our algorithm's time and space complexity are $O(n)$ and $O(U)$ respectively, and it works in an online manner. [1]

The $k$ maximum sum segments problem was introduced by Bae and Takaoka [5]. A natural extension of this problem is the $k$ length-constrained maximal density segments problem. The problem is defined as follows:

**Definition 3.** *Given a sequence of real numbers $a_i$, $i = 1, 2, \ldots, n$, a pair of real numbers $L \leqslant U$ and an integer $k$ such that $1 \leqslant k \leqslant (n - U + 1)(U - L + 1) + \frac{1}{2}(U - L)(U - L + 1)$. The $k$ length-constrained maximum density segments problem is to find $k$ subsequence of consecutive elements of length at least $L$ and at most $U$ such that their densities are the $k$ largest among all the possible segments of length at least $L$ and at most $U$.*

Combining with the technique of Brodal and Jorgensen [9] we extend our algorithm for the maximum density segment problem to solve the $k$ length-constrained (i.e., arbitrary $L$ and $U$) maximum density segments problem. Its time and space complexities are $O(n + k)$ and $O(U + k)$ respectively. Previously, there was no known algorithm with non-trivial result for this problem.

Huang [25] introduced the problem of finding segments of a sequence satisfying a sum requirement. The content requirement is expressed as the count of equal length oligomers in biomolecular sequence. We shall call this problem as the required sum segments problem. A natual extension of this is the required density segments problem. The problems are defined as:

**Definition 4.** *Given a sequence of real numbers $a_i$, $i = 1, 2, \ldots, n$, a real number $d$, and another pair of real numbers $L \leqslant U$. (a) The required sum segments problemm is to find all the subsequences $a_i, a_{i+1}, \ldots, a_j$ of length at least $L$ and at most $U$ such that $a_i + a_{i+1} + \ldots + a_j > d$. (b) The required density segments problemm is to find all the subsequences $a_i, a_{i+1}, \ldots, a_j$ of length at least $L$ and at most $U$ such that $(a_i + a_{i+1} + \ldots + a_j)/(w_i + w_{i+1} + \ldots + w_j) > d$.*

For the required sum segments problem when there is only lower bound on the length of the sequence and no upper bond on its length Huang [25] gave a linear time algorithm for the problem using dynamic programming technique. We describe an extension of our algorithm for the maximum sum segment problem to solve this problem for the general case, i.e., when $L$ and $U$ are arbitrary, in linear time. Previously, there was no known algorithm with non-trivial result for this case.

---

[1] A talk was given on the algorithm in the 20th Annual Fall Workshop on Computational Geometry 2010 [2].

Combining with the technique of Brodal and Jorgensen [9] we extend our algorithms for the maximum sum segment problem and the maximum density segment problem to solve respectively the required sum segments problem and the required density segments problem. Their time complexities are $O(\max\{n, m\})$ and $O(\max\{n \log L, m\})$ respectively, where $m$ is the number of output. Previously, there was no known algorithm with non-trivial result for these problems.

All of our algorithms can be used to solve the corresponding higher dimensional problems by reducing them to 1-dimensional problem in the way described by [7, 38]. They can also be extended to solve the problems with non-uniform width and non-uniform weight. We note that for $k$ maximum sum segments problem there is another version of the problem where there is no restriction on the segment length (i.e., $L = 0$ and $U = n$) but the segments are not allowed to overlap. For this case there are linear time algorithms for 1-dimension [10, 31]. In this paper, we shall not pursue this line. In all of our algorithms in this paper the segments are allowed to overlap.

According to [29, 36] the compositional heterogeneity of a genomic sequence is strongly correlated to its CG content regardless of the size of the genome. It is also found that gene length[14], gene density[41], patterns of codon usage[32], distribution of different types of repetitive elements[14, 35], number of isochores[8], length of isochore[29] and recombination rate within chrosome[19] are related to CG content. The algorithms can be used directly to find length constrained CG-rich regions with the maximum sum and average or with some user specified content requirement in a DNA sequence.

The nucleotide composition of a newly determined DNA sequence is analyzed to locate its biologically meaningful segments including finding CG-rich regions [15, 24], TA and CG-deficient regions [30], CpG islands [24], regions rich in periodical three-base pattern [33, 39], post processing sequence alignment [40], annotating multiple sequence alignments [36] and computing length constrained ungapped local alignment [3]. Our algorithms have potential applications in those areas.

In Section 2 we briefly describe Kim's [27] algorithm for the maximum density segment problem. Our algorithms for the maximum density and sum segments of a sequence are presented in Sections 3 and 4 respectively. Section 5 describes our algorithms for finding all the segments of a sequence having sum or density of at least some user specified value. Concluding remarks are given in Section 6.

## 2   Kim's Algorithm for Maximum Density Segment

We describe Kim's [27] algorithm for the maximum density segment problem using uniform width. He reduced the problem to a geometric one thus. Let $P[i] = a_1 + a_2 + \ldots + a_i$, be the $i$th prefix-sum of the given sequence $S : a_1, a_2, \ldots, a_n$ (define P[0]=0). This gives $n + 1$ points in the plane $p_0 = (0, P[0]), p_1 = (1, P[1]), p_2 = (2, P[2]), \ldots, p_n = (n, P[n])$, sorted by their $x$-coordinates. The density of a segment $a_i, a_{i+1}, \ldots, a_j$ can then be interpreted as the slope of the line segment through the points $(i-1, P[i-1])$ and $(j, P[j])$. The problem then is to find $p_i$ and $p_j$ such that $\overline{p_i p_j}$ has the largest slope.

Without any restriction on the segment length, the maximum density problem is solved by computing the largest slope defined by a pair of the above points. We can use any of a number of $O(n \log n)$ slope selection algorithms for this problem ([12] or [26] for example). The constraints on the segment length add a twist to the problem.

For a given *right* endpoint $p_j$, the set of candidate *left* endpoints $p_i$ has $i$ in the index-window $I_j = [0, j-L+1]$ when $L \leqslant j < U$ and in $I_j = [j-U+1, j-L+1]$ when $j \geq U$. If we maintain the lower convex hull of the points in this index-window, then the largest slope is found by drawing a tangent from $p_j$ to a point $p_t$ on this lower hull. The maximum density segment for a fixed $j$ is then $a_{t+1}, \ldots, a_j$. As $j$ goes from $l$ to $n$ the maximum of all slopes found gives the desired maximum density segment.

Based on the above formulation, Kim proposed an algorithm that claimed to be able to perform all the dynamic updates to the lower convex hull as the index-window moves from the left to the right in $O(n)$ amortized time. This claim is marred by the following problem. Figure 1 shows the lower convex hull (*lch*, for short) of the points inside the index-window $I(j)$, where $p_x, p_z$ and $p_y$ are the leftmost, bottommost and rightmost points on the *lch*. Kim maintains the portion of *lch* from $p_y$ to $p_z$ in one array and the portion of the *lch* from $p_x$ to $p_z$ in another array.

Now, it is crucial to the correctness of Kim's algorithm that, as the window $I_j$ slides to the right the algorithm remains updated about the new value of $p_z$. Kim's algorithm correctly updates $p_z$, except in the case shown in Figure 1(b).

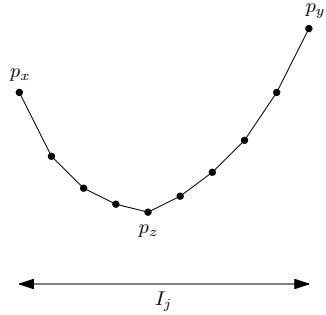

**Fig. 1.** *The lower convex hull of the points in the index-window $I_j$*

In this case, as the window slides to the next position the hull update cannot be done in $O(1)$ time as Figure 1(c) shows.

The blind-spot in Kim's algorithm is that as the index-window $I_j$ slides to the right, the lowest point $p_z$ under the tacit assumption that the situation as described above never arises. It can as shown above.

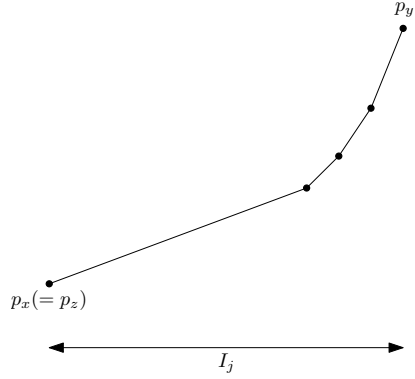

**Fig. 2.** *The problem case for Kim's algorithm*

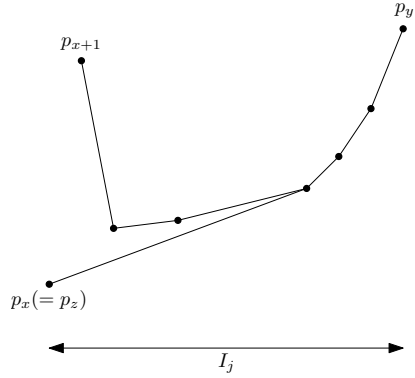

**Fig. 3.** $p_z$ *may need to be recomputed*

## 3   Our Algorithm for Maximum Density Segment

First, we describe our algorithm for the case of uniform width, i.e., $w_i = 1$ for $i = 1, ..., n$. The main idea underlying the new algorithm is to consider the right end point $p_j$ (for $j = L, L+1, ..., n$) of a candidate largest slope segment $\overline{p_i p_j}$ in batches of a fixed size. For each $p_j$, instead of computing a single lower convex hull of the feasible set of left points $p_i$, we compute two lower convex hulls - a left one and a right one that are joined at a common extreme point $p_k$, $j - U + 1 \leqslant k \leqslant j - L + 1$ (Fig. 4). The right lower hulls are computed incrementally in a left-to-right ($LR$) pass for the batched set $p_j$, and the left hulls in a right-to-left ($RL$) pass for the same batched set. Thus the problem that arises in Kim's algorithm from the dynamic convex hull update as a result of deletion on the left is avoided. The correctness of this scheme is due to the following easily-proved lemma.

**Lemma 1.** *For a point $p_j, L \leqslant j \leqslant n$, let the candidate left end points $p_i$, $i \in [j - U + 1, j - L + 1]$ of all feasible segments be divided into two groups:*

8  Md. Shafiul Alam and Asish Mukhopadhyay

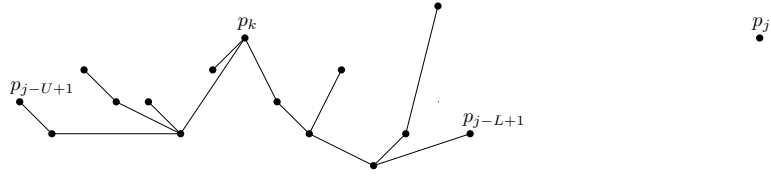

**Fig. 4.** *Incremental left and right lower convex hulls.*

$G_1 = \{p_i | i \in [j - U + 1, j - U + k]\}$ and $G_2 = \{p_i | i \in [j - U + k, j - L + 1]\}$, with $1 \leqslant k \leqslant U - L + 1$. Then the maximum slope of a segment $\overline{p_i p_j}$ with $p_i \in G_1 \cup G_2$ is the maximum of two maximum slopes obtained by restricting $p_i$ to be first in $G_1$ and then in $G_2$.

The right end points $p_j$ are considered in batches of size $U - L + 1$, where $j \in [k, k + U - L]$ and $k \geq U$. The details of the $LR$ and $RL$ passes are as follows.

### 3.1  The LR pass

We incrementally compute the convex hulls $CH(\{p_{k-L+1}, p_{k+1-L+1}, \ldots, p_{j-L+1)}\})$, for $j = k, \ldots, k + U - L$. Following Kim [27], we maintain 3 parameters to aid the incremental computation : the maximum slope $\mu$ found so far, a tangent line $l$ to the current hull with slope $\mu$ and the point of contact $\alpha$ of $l$ with the current hull.

Initially, $l = \overline{p_{k-L+1} p_k}$, $\mu = slope(l)$ and $\alpha = p_{k-L+1}$. For $j = k + t$, where $0 \leqslant t \leqslant U - L$, we update the right lower hull, $\mu$, $l$ and $\alpha$ according to the 4 cases below:

**Case 1** $p_{k-L+1+t}$ and $p_{k+t}$ are both above $l$ (Fig 5).
  The (right) hull is updated. $\alpha$ is set to the point of contact of the tangent from $p_{k+t}$ to this new hull, while this tangent and its slope are set to be the new $l$ and $\mu$ respectively.
**Case 2** $p_{k-L+1+t}$ is above $l$ and $p_{k+t}$ is below $l$ (Fig 6).
  The (right) hull is updated. However $\mu$, $\alpha$ and $l$ remain unchanged.
**Case 3** $p_{k-L+1+t}$ is below $l$ (Fig 7).
  The (right) hull is updated. Let $l'$ be a line through $p_{k-L+1+t}$ parallel to $l$. Let $p_{k+t}$ be above $l'$; reset $l = \overline{p_{k-L+1+t} p_{k+t}}$; $\mu = \text{slope}(l)$ and $\alpha = p_{k-L+1+t}$.
**Case 4** $p_{k-L+1+t}$ is below $l$ and $p_{k+t}$ is below $l'$ (Fig 8).
  The (right) hull is updated. Set $l$ to $l'$ and $\alpha = p_{k-L+1+t}$

As $p_{k-L+1+t}$ and hence $p_{k+t}$ both move right $\alpha$ never moves to the left. So, the cost for this pass is linear in the number of $p_j$'s considered.

### 3.2  The RL pass

This pass needs more careful handling. We incrementally compute the convex hulls $CH(\{p_{k-L+1}, p_{k+1-L-1}, \ldots, p_{j-U+1}\})$ for $j = k + U - L \ldots k$.

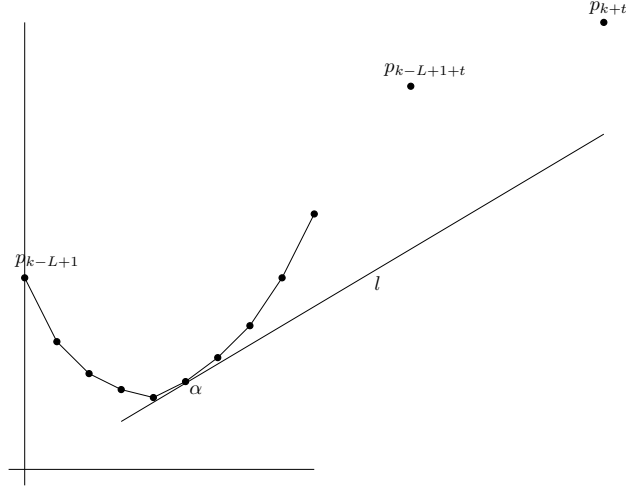

**Fig. 5.** Both $p_{k-L+1+t}$ and $p_{k+t}$ are above $l$

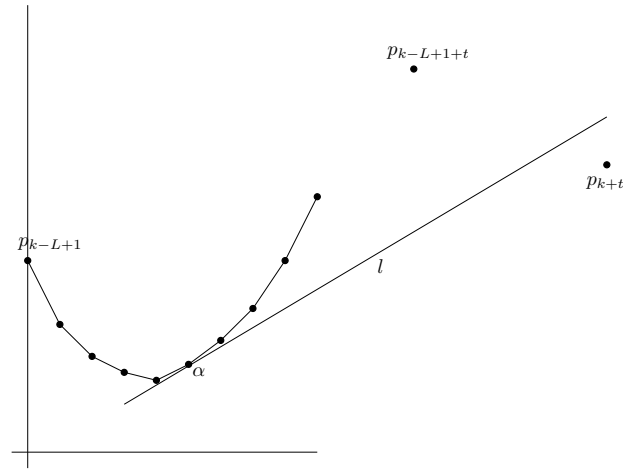

**Fig. 6.** $p_{k-L+1+t}$ is above while $p_{k+t}$ is below $l$

Initially, $l = \overline{p_{k-L+1}p_{k+U-L}}$, $\mu = slope(l)$ and $\alpha = p_{k-L+1}$. For $j = k+U-L-t$, where $0 \leqslant t \leqslant U-L$, we update the left lower hull according to the 4 cases below:

**Case** 1 $p_{k-L+1-t}$ and $p_{k+U-L-t}$ are both above $l$ (Fig. 9).
  The (left) hull is updated. In this case the update does not go left beyond $\alpha$ on the current hull. We traverse the convex hull counterclockwise from $\alpha$ to check if a tangent can be drawn to it from $p_{k+U-L-t}$ that has a larger slope than $\mu$. If so we reset $\alpha$, $\mu$ and $l$ appropriately.

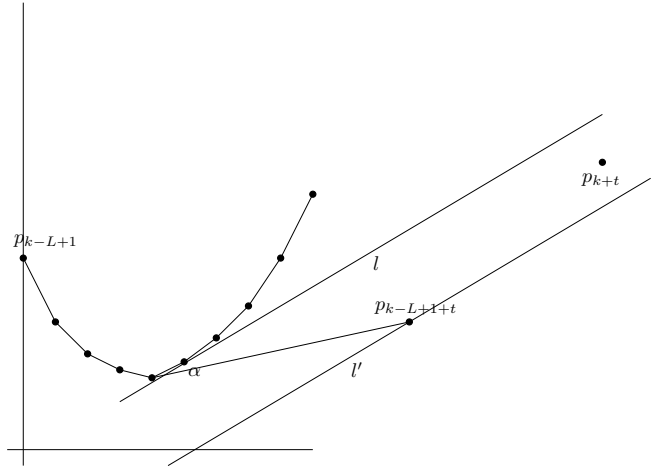

**Fig. 7.** $p_{k-L+1+t}$ is below $l$ and $p_{k+t}$ is above $l'$

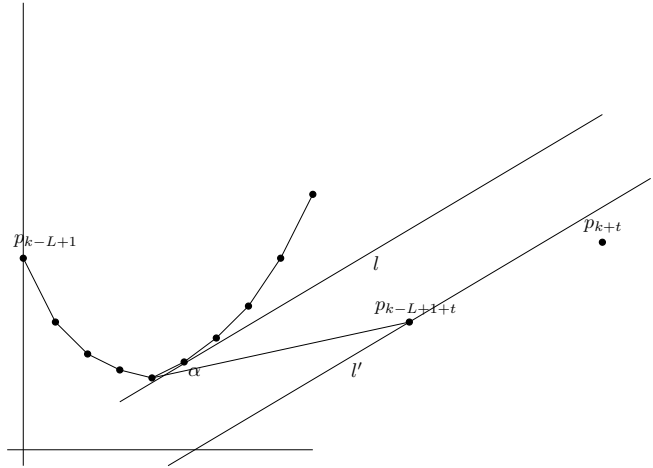

**Fig. 8.** $p_{k-L+1+t}$ is below $l$ and $p_{k+t}$ is below $l'$

**Case** 2 $p_{k-L+1-t}$ is above $l$ and $p_{k+U-L-t}$ is below $l$ (Fig. 10).
  The (left) hull is updated. However $\mu$, $\alpha$ and $l$ remain unchanged.
**Case** 3 $p_{k-L+1-t}$ is below $l$ and $p_{k+U-L-t}$ is above $l$ (Fig. 11).
  The (left) hull is updated to a vertex beyond $\alpha$. We traverse the updated hull from the newly added point counterclockwise to check if a a tangent can be drawn to it from $p_{k+U-L-t}$ that has a larger slope than $\mu$. If so we reset $\alpha$, $\mu$ and $l$ appropriately. In this case, $\alpha$ can move left to the newly added point.
**Case** 4 $p_{k-L+1-t}$ is below $l$ and $p_{k+U-L-t}$ is below $l'$ (Fig. 12).
  The (left) hull is updated as in Case 3. We check if the join of $p_{k-L+1-t}$ and $p_{k+U-L-t}$ has a larger slope than $\mu$ (this is easily done by discriminating with respect to a line through $p_{k-L+1-t}$ parallel to $l$). In that case we reset $\alpha$ to $p_{k-L+1-t}$, $l$ to $\overline{p_{k-L+1-t}p_{k+U-L-t}}$ and $\mu$ to the slope of the new $l$.

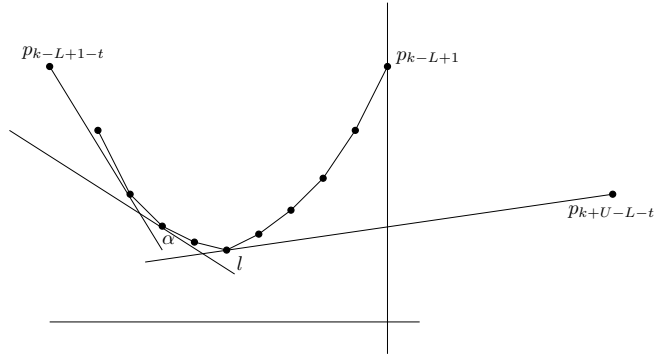

**Fig. 9.** $p_{k-L+1-t}$ *is above* $l$ *and* $p_{k+U-L-t}$ *is above* $l$

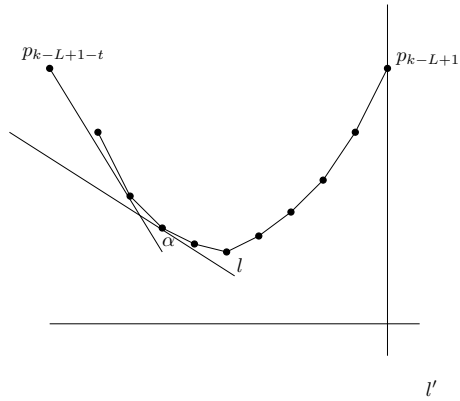

**Fig. 10.** $p_{k-L+1-t}$ *is above* $l$ *and* $p_{k+U-L-t}$ *is below* $l$

As both $p_{k+U-L-t}$ and $p_{k+L-1-t}$ moves left $\alpha$ may move backward to the left at most once for each $p_j$. So, the cost for this pass is linear in the number of $p_j$'s considered.

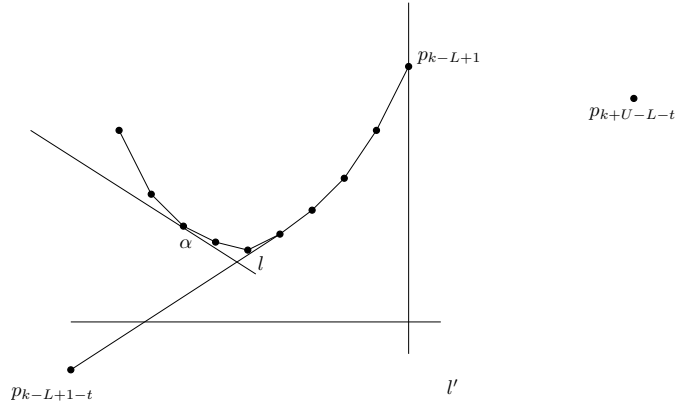

**Fig. 11.** $p_{k-L+1-t}$ is below $l$ and $p_{k+U-L-t}$ is above $l$

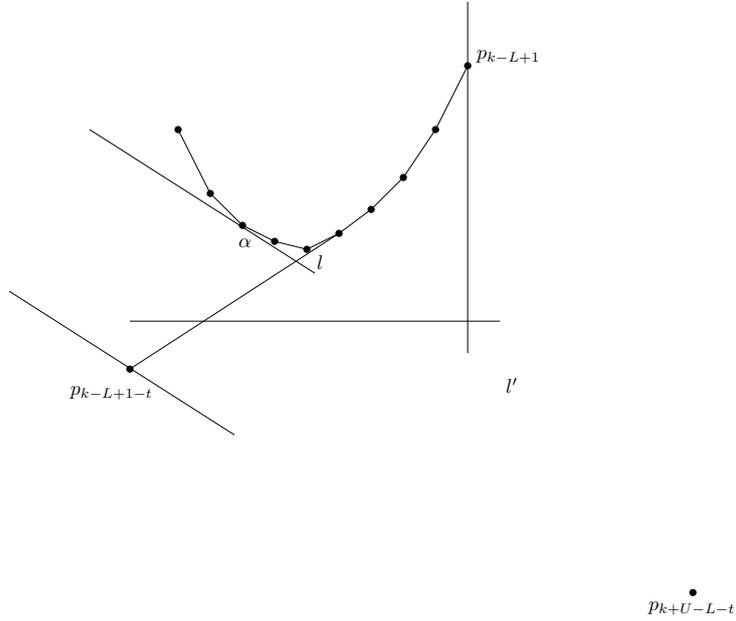

**Fig. 12.** $p_{k-L+1-t}$ is below $l$ and $p_{k+U-L-t}$ is also below $l$

### 3.3 Analysis

Initially, we find the maximum density for the segments with right end points $p_j, j = L, L+1, ..., U$ using the incremental lower convex hull in a right pass as described above. Then for each batch of right end points $p_j, j = U + b(U - L), U + b(U - L) + 1, ..., U + (b+1)(U - L)$, where $b$ is an integer with values $b = 1, ..., \lfloor (n - U)/(U - L) \rfloor$, we find the maximum density for all the feasible segments having these points as right end points in two passes as described above. For the residual set of the $p_j$'s we use an $LR$ pass only. The indices of the subsequence whose density has the maximum slope from all these passes is returned as the result. It is clear from the way the computation has been organized that the time complexity is in $O(n)$. Thus, we have the following theorem:

**Theorem 1.** *Given a sequence A of n real numbers and two real numbers L and U with $1 \leqslant L \leqslant U \leqslant n$, our geometric algorithm as described above finds the maximum density segment of A from among all the segments of A of length at least L and at most U in linear time in an online manner.*

The above algorithm for uniform widths can be extended to solve the general problem where the widths, $U$ and $L$ are arbitrary. For this we define the cumulative width $W_i, i = 0, ..., n$, as $W_0 = 0$ and $W_i = w_1 + ... + w_i$. Then the density $\mu_{i,j}$ of a segment $S(i,j)$ can be written as

$$\mu_{i,j} = \frac{P_j - P_{i-1}}{W_j - W_{i-1}}.$$

For each element $a_i, i = 1, ..., n$, in the sequence $S$ we get the point $(P[i], W[i])$, $i = 1, ..., n$, in the plane. We have the initial point $(P[0], W[0]) = (0, 0)$. Then the problem to find the feasible segments with the maximum sum is reduced to finding the feasible pairs of points with the maximum slope. And this can be solved by a simple modification to our above algorithm for uniform widths. The only difference is that the abscissas of the consecutive points $P[i], W[i]$ and $P[i+1], W[i+1]$ are $w_{i+1}$ distance away instead of equal distance away. Its time and space complexity will remain $O(n)$ and $O(U)$ as before. Thus, we have the following theorem:

**Theorem 2.** *Given a sequence A of n pairs of real numbers $(a_i, w_i), i = 1, ..., n$, and two real numbers L and U with $1 \leqslant L \leqslant U \leqslant n$, our geometric algorithm as described above finds the maximum density segment of A from among all the segments of A of length at least L and at most U in linear time in an online manner.*

Using the method of [7, 38] the 2-dimensional problem is reduced to $\binom{m}{2} + m$ number of 1-dimensional problems, where the input is an $m \times n$ matrix. We solve each of them using the above algorithm. The time complexity will be $O(m^2 n)$.

**Theorem 3.** *Given a 2-dimensional $m \times n$ matrix A of pairs of real numbers $(a_{ij} w_{ij}), i = 1, ..., m; j = 1, ..., n$, and two real numbers L and U with $1 \leqslant L \leqslant U \leqslant n$, there exists an algorithm to find the maximum density subarray of A from among all the subarrays of A of length at least L and at most U in $O(m^2 n)$ time and $O(mU)$ space.*

*Proof.* Omitted.

The above algorithm can be extended to any dimension $d$ in a straight forward way.

**Theorem 4.** *Given a d-dimensional $n_1 \times n_2 \times ... \times n_d$ matrix A of pairs of real numbers and two real numbers L and U with $1 \leqslant L \leqslant U \leqslant n$, there exists an algorithm to find the maximum density subarray of A from among all the subarrays of A of length at least L and at most U in $O(n_1 \Pi_{i=2}^{d} n_i^2)$ time and $O(U \Pi_{i=2}^{d} n_i)$ space.*

*Proof.* Omitted.

All of our remaining algorithms in this paper can be extended in a similar way to solve the corresponding higher dimensional problems by using the technique of reducing the problems to 1-dimension [7, 38].

### 3.4  *k* Maximum Density Segments

In this paper, we shall use the term heap to denote maximum heap ordered binary tree. First, we consider the case $k \leqslant U - L$. The above SPLITHULL algorithm is repeated $k$ times for each batch of $U - L + 1$ points to find at least $k$ maximum density segments with right end points in the batch. In each iteration the maximum density segment for the iteration is found. Keeping the left end point of this segment fixed all the valid segments with right end point within the current batch of $U - L + 1$ points are selected. Then the left end point is deleted at the start of the next iteration. All these maximum density segments are inserted into a heap $H$ [34]. The $k$ largest density segments are selected from $H$ in $O(k)$ times using the binary heap selection algorithm of Frederickson [16]. The left pass of the algorithm for a batch of $U - L + 1$ points is described below.

---

Algorithm **MDS-SMALLK**
    1.    Let $X$ and $Y$ be the sets of $U - L + 1$ number of left and right end points of algorithm SPLITHULL.
2.    For $i = 1$ to $k$ do.
2.1.    Find the maximum density segment from among the valid segments with left end points in $X$ and right end points in $Y$ using the left pass of the SPLITHULL algorithm.
2.2.    Let $x$ and $y$ be the left and right end points of this segment respectively. Insert $(x, y, d(x, y))$ in the skew heap $H$.
2.3.    For all valid segments with left end point $x$, i.e., $(x, y'), y' \in Y$, insert $(x, y', d(x, y'))$ in the skew heap $H$.
2.4.    Delete $x$ from $X$.

---

In each iteration the algorithm will select at most $U - L + 1$ segments. In $k$ iterations it will select at most $k(U - L + 1)$ segments. The right pass of the algorithm for the same batch of points is similar. It uses the right pass of SPLITHULL. Thus the maximum number of points selected for the batch of $U - L + 1$ right end points is $2k(U - L + 1)$. The cost of each insertion in $H$ is in $O(1)$. For the batch of $U - L + 1$ points, the total cost of selection of maximum density segments and inserting them in $H$ is $O(k(U − L + 1))$. Thus the total cost per right end point is in $O(k)$.

**Theorem 5.** *Given a sequence $A$ of $n$ real numbers, two integers $L$ and $U$ with $1 \leqslant L \leqslant U \leqslant n$, and one integer $k \leqslant U − L$, MDS-SMALLK algorithm finds the $k$ maximum density segments of $A$ from among all the segments of $A$ of length at least $L$ and at most $U$ in $O(kn)$ time in online manner.*

When $k > U − L$ we do not use the SPLITHULL algorithm. Instead, for each of $\lceil \frac{k}{U−L+1} \rceil$ number of batches of $U − L + 1$ consecutive elements of the sequence we insert into $H$ all the possible segments with right end elements in those batches. Select $2k$ maximum segments from $H$, using Frederickson's heap selection algorithm. Then construct $H$ using these segments. The total cost is clearly $O(U − L)$ for each right end point. At the end of processing of the last right end point we have found $2k$ maximum density segments. From these, we find the $k$-th maximum in $O(k)$ time. Thus, we have the following theorem:

**Theorem 6.** *Given a sequence $A$ of $n$ real numbers, two integers $L$ and $U$ with $1 \leqslant L \leqslant U \leqslant n$, and one integer $k > U − L$, the above algorithm finds the $k$ maximum density segments of $A$ from among all the segments of $A$ of length at least $L$ and at most $U$ in $O(n(U − L))$ time in an online manner.*

## 4   Maximum Sum Segment

We shall use a simple modification of Brodal and Jorgensen [9] method to solve the maximum sum problem. As before we solve the problem in batch mode with $U − L$ number of elements in each batch as right end elements of the feasible segments being considered in a batch. Analogous to Lemma 1 for the maximum density segment problem we have the following trivial lemma:

**Lemma 2.** *For an element $a_j, L \leqslant j \leqslant n$, let the candidate left end elements $a_i, i \in [j − U + 1, j − L + 1]$ of all feasible segments be divided into two groups: $G_1 = \{a_i | i \in [j − U + 1, j − U + k]\}$ and $G_2 = \{a_i | i \in [j − U + k, j − L + 1]\}$, with $1 \leqslant k \leqslant U − L + 1$. Then the maximum sum of all the feasible segments $S(i, j)$ with $a_i \in G_1 \cup G_2$ is the maximum of the two maximum sums obtained by restricting $a_i$ to be first in $G_1$ and then in $G_2$.*

So, for each batch of $U − L$ right end points we make 2 passes as before. For a pass we shall use Brodal and Jorgensen [9] algorithm to construct a partially persistent [13] max-heap that implicitly contains all the feasible segments with their respective sums. It will take $O(U)$ time and space to build it. From this heap we select the maximum element in constant time. For each pass of each batch of $U − L$ right end elements we update the maximum. To build the heap we note that the sets of all feasible segments have right end points $a_j$, where $a_j$ is one of the right end elements in the current batch of right end elements. Let the batch of elements be $a_j, ..., a_{j+U−L}$, where $j = k(U − L), k = 1, ..., \frac{n}{U−L}, j \geq U$. Here we describe the RL pass. The LR pass is similar and simpler. The incremental construction of the heap elements $(\delta_{k+1}, \delta'_{k+1}, H_{suf}^{k+1})$ are shown below:

$$(\delta_{j+1}, \delta'_{j+1}, H_{suf}^{j+1}) = (S(j - U + L + 1, j - L + 1), S(j - U + L + 1), \phi),$$

$$(\delta_{m+1}, \delta'_{m+1}, H_{suf}^{m+1}) = (\delta_m + a_{m+1}, \delta'_m - a_{j-m}, H_{suf}^m \cup \{\delta'_m - \delta_m\}).$$

where $m = j, j - 1, ..., j - U + L - 1$. Thus, we have the following theorem:

**Theorem 7.** *Given a sequence A of n real numbers and two real numbers L and U with $1 \leqslant L \leqslant U \leqslant n$, our algorithm as described above finds the maximum sum segment of A from among all the segments of A of length at least L and at most U in linear time in online manner.*

*Proof.* Omitted.

By making a simple modification as was done for our algorithm for the maximum density segment problem for non-uniform case, the ALL-HEAVIER-SEGMENTS algorithm can solve the problem for the non-uniform case in the same time and space complexity.

### 4.1  *k* Maximum Sum Segments

To find the $k$ maximum sum segments we extend the above algorithm according to Brodal and Jorgensen [9]. As before we work in batch mode with batch of feasible segments with $U - L + 1$ right end elements. For all the feasible segments in the batch we construct the partially persistent max-heap as before. We select the largest $k$ elements from this heap in $O(k)$ time. We insert each of them into the self adjusting (skew) heap $H^{i-1}$ of Sleator and Tarjan [34] in amortized constant time. We select the largest $k$ elements from this heap using Frederickson's [16] heap selection algorithm and insert them into a new skew heap $H^i$. We delete the old skew heap $H^{i-1}$. Thus, we have the following theorem:

**Theorem 8.** *Given a sequence A of n real numbers, two integers L and U with $1 \leqslant L \leqslant U \leqslant n$, and one integer $k \leqslant U - L$, there exists an algorithm to find the k maximum sum segments of A from among all the segments of A of length at least L and at most U in $O(n)$ time and $O(U)$ space.*

When $k \geq U - L$ we use the above algorithm for each of $\lceil \frac{k}{U-L+1} \rceil$ number of batches of $U - L + 1$ consecutive elements of the sequence we insert into $H$ all the possible segments with right end points in that batch. Select the $2k$ maximum segment from the heap using Frederickson's heap selection algorithm. Then construct $H$ using these segments. The total cost is clearly $O(U - L)$ for each right end point. At the end of processing of the last right end point we have found $2k$ maximum density segments. From them we find the $k$-th maximum in $O(k)$ time. Thus, we have the following theorem:

**Theorem 9.** *Given a sequence A of n real numbers, two real numbers L and U with $1 \leqslant L \leqslant U \leqslant n$, and one integer $k > U - L$, there exists an algorithm to find the k maximum sum segments of A from among all the segments of A of length at least L and at most U in $O(n + k)$ time and $O(k)$ space.*

## 5 Finding All the Segments with Some Content Requirement

In genomic sequence analysis sometimes it is necessary to find all the segments of a sequence with some user specified minimum sum or density requirements[25]. In this section we first describe the case of sum and then the case of density.

### 5.1 Finding All the Segments with a Minimum Sum

We construct the partially persistent self adjusting binary heap as before. From this heap we select all the segments having sum of at least the user specified amount.

### 5.2 Finding All the Segments with a Minimum Density

As before our algorithm work in batch mode. For each batch of $U - L + 1$ number of right end points we find all the segments with left end points such that the density of each of them is at least the user specified minimum density. This is done in two passes as before. For each pass, the set of left end points of SPLITHULL algorithm is sorted incrementally with the new left end point corresponding to the new right end point being put into its sorted position. The relative positions of a pair of points are determined by the distance between the pair of parallel straight lines passing through them and having slope of user specified minimum density. For the new right end point the corresponding set of sorted left end points are searched to find a segment with density at least that of the given amount. Then all the segments with densities higher than that are selected. The left pass of the algorithm for a batch of $U - L + 1$ points is described below. In the algorithm $X$ and $Y$ are respectively the sets of $U - L + 1$ number of left and right end points of algorithm SPLITHULL and $d$ is the user specified density requirement.

---

Algorithm **ALL-HEAVIER-SEGMENTS**($X,Y,d$)
1.     Let $X' = \phi$ and $S = \phi$.
2.     For $i = 1$ to $U - L + 1$ do.
2.1.        Let $x_i \in X$ and $y_i \in Y$ be the new left and right end points respectively. Insert $x_i$ into $X'$ in its sorted position $i'$ according to the measure of relative position and the method of comparison described above. Let $X$ be sorted in the increasing order of $x$-intercept of the line passing through the left end point and having slope of the user specified minimum density.

2.2.     Search $X'$ using binary search method to find the element $x_{i'} \in X'$ for which the segment $(x_{i'}, y_i)$ has density at least d. Let $H$.
2.3.     From $X'$ select all the elements $x'_j$ in the part from $x_{i'}$ to the end of $X'$. Insert $(x'_j, y_i)$ in $S$ for all such $x'_j$.
3.     Return $S$.

The right pass of the algorithm for the same batch of points is similar. Sorting the set of left end elements take $O((U-L)\log(U-L))$ time. For each right end point the binary search to find the left end point in the set $X'$ take a maximum of $O(\log(U-L))$ time. Thus, the total time is $O(\max(n\log(U-L), m))$, where $m$ is the number of output.

**Theorem 10.** *Given a sequence $A$ of $n$ real numbers, two integers $L$ and $U$ with $1 \leqslant L \leqslant U \leqslant n$, and one real number d, ALL-HEAVIER-SEGMENTS algorithm finds all the segments of $A$ of length at least $L$ and at most $U$ having density at least d in $O(\max(n\log(U-L), m))$ time in online manner, where $m$ is the number of output.*

By making the same modification as was done for our algorithm for the maximum density segment problem for non-uniform case, the ALL-HEAVIER-SEGMENTS algorithm can solve the problem for the non-uniform case in the same time and space complexity.

## 6   Conclusions

In this paper two problems concerning the search for the interesting regions in a sequence are considered. The problems are to find a consecutive subsequence of length at least $L$ and at most $U$ with the maximum sum and density respectively. We have presented linear time algorithms for both the problems. We have extended our algorithms to find the $k$ segments of length at least $L$ and at most $U$ with the largest sum and density. We have also extended our algorithms to find all the segments with user specified sum or density. We indicate the extensions of our algorithms to higher dimensions. Our algorithms facilitate efficient solutions for all these problems in higher dimensions. All the algorithms can be extended in a straight forward way to solve the problems with non-uniform width and non-uniform weight.

The algorithms have applications in several areas of biomolecular sequence analysis including finding CG-rich regions, TA and CG-deficient regions, regions rich in periodical three-base pattern, post processing sequence alignment, annotating multiple sequence alignments and computing length constrained ungapped local alignment.

It would be interesting to study if there is any linear time algorithm to find the $k$-th density segment with length between the lower and upper bounds $L$ and $U$ respectively. It is also interesting to investigate if there is any linear time algorithm in the number of input and output to find all the segments with length between $L$ and $U$ and satisfying user specified minimum density requirement.

It can also be investigated to find more efficient algorithms for the problems in higher dimensions. It remains open to improve the trivial lower bounds for these cases.

# References

1. R. Agrawal, T. Imieliński, and A. Swami. Mining association rules between sets of items in large databases. *SIGMOD Rec.*, 22:207–216, June 1993.
2. M. S. Alam and A. Mukhopadhyay. A new geometric algorithm for the maximum density segment problem. In *Proceedings of the 20th Annual Fall Workshop on Computational Geometry*, 2010.
3. N. N. Alexandrov and V. V. Solovyev. Statistical significance of ungapped alignments. In *Pacific Symposium on Biocomputing (PSB-98)*, pages 463–472. 1998.
4. L. Allison. Longest biased interval and longest non-negative sum interval. *Bioinformatics*, 19(10):1294–1295, 2003.
5. S. E. Bae and T. Takaoka. Algorithms for the problem of k maximum sums and a vlsi algorithm for the k maximum subarrays problem. *Parallel Architectures, Algorithms, and Networks, International Symposium on*, 0:247, 2004.
6. J. Bentley. Programming pearls: algorithm design techniques. *Commun. ACM*, 27:865–873, September 1984.
7. J. Bentley. Programming pearls: perspective on performance. *Commun. ACM*, 27:1087–1092, November 1984.
8. G. Bernardi. Isochores and the evolutionary genomics of vertebrates. *Gene*, 241(1):3 – 17, 2000.
9. G. S. Brodal and A. G. Jorgensen. A linear time algorithm for the $k$ maximal sums problem. In *Proc. 32nd International Symposium on Mathematical Foundations of Computer Science*, volume 4708 of *Lecture Notes in Computer Science*, pages 442–453. Springer Verlag, Berlin, 2007.
10. K.-Y. Chen and K.-M. Chao. On the range maximum-sum segment query problem. In R. Fleischer and G. Trippen, editors, *Algorithms and Computation*, volume 3341 of *Lecture Notes in Computer Science*, pages 201–210. Springer Berlin / Heidelberg, 2005.
11. K.-M. Chung and H.-I. Lu. An optimal algorithm for the maximum-density segment problem. *SIAM J. Comput.*, 34(2):373–387, 2005.
12. R. Cole, J. S. Salowe, W. L. Steiger, and E. Szemeredi. An optimal-time algorithm for slope selection. *SIAM J. Comput.*, 18(4):792–810, 1989.
13. J. R. Driscoll, N. Sarnak, D. D. Sleator, and R. E. Tarjan. Making data structures persistent. *Journal of Computer and System Sciences*, 38(1):86 – 124, 1989.
14. L. Duret, D. Mouchiroud, and C. Gautier. Statistical analysis of vertebrate sequences reveals that long genes are scarce in gc-rich isochores. *Journal of Molecular Evolution*, 40:308–317, 1995. 10.1007/BF00163235.
15. C. Fields and C. Soderlund. gm: a practical tool for automating dna sequence analysis. *Computer applications in the biosciences : CABIOS*, 6(3):263–270, 1990.
16. G. N. Frederickson. An optimal algorithm for selection in a min-heap. *Information and Computation.*, 104(2):197–214, 1993.
17. T. Fukuda, Y. Morimoto, S. Morishita, and T. Tokuyama. Data mining using two-dimensional optimized association rules: sche algorithms, and visualization. *SIGMOD Rec.*, 25:13–23, June 1996.

18. T. Fukuda, Y. Morimoto, S. Morishita, and T. Tokuyama. Data mining with optimized two-dimensional association rules. *ACM Trans. Database Syst.*, 26:179–213, June 2001.
19. S. M. Fullerton, A. Bernardo Carvalho, and A. G. Clark. Local rates of recombination are positively correlated with gc content in the human genome. *Molecular Biology and Evolution*, 18(6):1139–1142, 2001.
20. M. Goldwasser, M.-Y. Kao, and H.-I. Lu. Fast algorithms for finding maximum-density segments of a sequence with applications to bioinformatics. In R. Guig and D. Gusfield, editors, *Algorithms in Bioinformatics*, volume 2452 of *Lecture Notes in Computer Science*, pages 157–171. Springer Berlin / Heidelberg, 2002.
21. M. H. Goldwasser, M.-Y. Kao, and H.-I. Lu. Linear-time algorithms for computing maximum-density sequence segments with bioinformatics applications. *Journal of Computer and System Sciences*, 70(2):128 – 144, 2005.
22. U. Grenander. *Pattern Analysis*. Springer-Verlag, New York, NY, USA, 1978.
23. D. Gries. A note on a standard strategy for developing loop invariants and loops. *Science of Computer Programming*, 2(3):207 – 214, 1982.
24. R. Hardison, D. Krane, D. Vandenbergh, J.-F. Cheng, J. Mansberger, J. Taddie, S. Schwartz, X. Huang, and W. Miller. Sequence and comparative analysis of the rabbit [alpha]-like globin gene cluster reveals a rapid mode of evolution in a g + c-rich region of mammalian genomes. *Journal of Molecular Biology*, 222(2):233 – 249, 1991.
25. X. Huang. An algorithm for identifying regions of a dna sequence that satisfy a content requirement. *Computer applications in the biosciences : CABIOS*, 10(3):219–225, 1994.
26. M. J. Katz and M. Sharir. Optimal slope selection via expanders. *Inf. Process. Lett.*, 47(3):115–122, 1993.
27. S. K. Kim. Linear-time algorithm for finding a maximum-density segment of a sequence. *Inf. Process. Lett.*, 86(6):339–342, 2003.
28. Y.-L. Lin, T. Jiang, and K.-M. Chao. Efficient algorithms for locating the length-constrained heaviest segments with applications to biomolecular sequence analysis. *Journal of Computer and System Sciences*, 65(3):570 – 586, 2002.
29. A. Nekrutenko and W. H. Li. Assessment of compositional heterogeneity within and between eukaryotic genomes. 10.
30. S. Ohno. Universal rule for coding sequence construction: Ta/cg deficiency-tg/ct excess. *Proceedings of the National Academy of Sciences*, 85(24):9630–9634, 1988.
31. W. L. Ruzzo and M. Tompa. A linear time algorithm for finding all maximal scoring subsequences. In *Proceedings of the seventh international conference on intelligent systems for molecular biology*, pages 234 – 241, 1999.
32. P. M. Sharp, M. Averof, A. T. Lloyd, G. Matassi, and J. F. Peden. DNA Sequence Evolution: The Sounds of Silence. *Royal Society of London Philosophical Transactions Series B*, 349:241–247, Sept. 1995.
33. J. C. Shepherd. Method to determine the reading frame of a protein from the purine/pyrimidine genome sequence and its possible evolutionary justification. *Proceedings of the National Academy of Sciences*, 78(3):1596–1600, 1981.
34. D. D. Sleator and R. E. Tarjan. Self adjusting heaps. *SIAM J. Comput.*, 15(1):52–69, 1986.
35. P. Soriano, M. Meunier-Rotival, and G. Bernardi. The distribution of interspersed repeats is nonuniform and conserved in the mouse and human genomes. *Proceedings of the National Academy of Sciences*, 80(7):1816–1820, 1983.

36. N. Stojanovic, L. Florea, C. Riemer, D. Gumucio, J. Slightom, M. Goodman, W. Miller, and R. Hardison. Comparison of five methods for finding conserved sequences in multiple alignments of gene regulatory regions. *Nucleic Acids Research*, 27(19):3899–3910, 1999.
37. T. Takaoka. Efficient algorithms for the maximum subarray problem by distance matrix multiplication. *Electronic Notes in Theoretical Computer Science*, 61:191 – 200, 2002. CATS'02, Computing: the Australasian Theory Symposium.
38. H. Tamaki and T. Tokuyama. Algorithms for the maximum subarray problem based on matrix multiplication. In *Proceedings of the ninth annual ACM-SIAM symposium on Discrete algorithms*, SODA '98, pages 446–452, Philadelphia, PA, USA, 1998. Society for Industrial and Applied Mathematics.
39. E. N. Trifonov. Translation framing code and frame-monitoring mechanism as suggested by the analysis of mrna and 16 s rrna nucleotide sequences. *Journal of Molecular Biology*, 194(4):643 – 652, 1987.
40. Z. Zhang, P. Berman, T. Wiehe, and W. Miller. Post-processing long pairwise alignments. *Bioinformatics*, 15(12):1012–1019, 1999.
41. S. Zoubak, O. Clay, and G. Bernardi. The gene distribution of the human genome. *Gene*, 174(1):95 – 102, 1996.